\documentclass[11pt]{article}
\usepackage[a4paper,margin=2.5cm]{geometry}
\usepackage{graphicx}
\usepackage[noend]{algpseudocode}
\usepackage{hyperref}
\usepackage{amsthm}
\usepackage{amsmath}
\usepackage{algorithm}
\usepackage{amssymb}
\usepackage{textcomp}
\usepackage{subfiles}
\usepackage{float}
\usepackage[capitalise]{cleveref}
    \crefname{observation}{Observation}{Observations}
    \Crefname{observation}{Observation}{Observations}
\usepackage{tikz} \usetikzlibrary{shapes.misc,decorations.pathreplacing}
\usepackage{xcolor}
\usepackage{bbm}
\usepackage{tabularx}
\usepackage{xspace}
\usepackage[backend=biber, style=numeric]{biblatex}
\newcommand{\predError}{\mathbf{B}}
\newcommand{\upError}{\tilde{\mathbf{B}}}
\newcommand{\preds}{\vec{a}}
\newcommand{\clas}{\vec{c}}
\newcommand{\inputdomain}{\{0,1\}}
\newcommand{\lead}{\textit{L}}
\newcommand{\king}{\texttt{king}}
\newcommand{\pos}{\texttt{pos}}

\newcommand{\Broadcast}{\textsc{Broadcast}}
\newcommand{\Receive}{\textsc{Receive}}
\newcommand{\msg}{\mathit{msg}}

\newcommand{\Honest}{\mathbf{H}}
\newcommand{\Faulty}{\mathbf{F}}
\newcommand{\predhonest}{\Honest^{\mathrm{pred}}}
\newcommand{\type}{\textsc{type}}
\newcommand{\Misclas}{\mathbf{M}}

\newcommand{\common}{j^{\ast}}
\newcommand{\goodcommon}{j_{\mathrm g}^{\ast}}
\newcommand{\nongoodcount}{d}
\newcommand{\candidateunion}{\mathcal C}
\newcommand{\densityset}{\mathcal D}
\newcommand{\sizeobjective}{\Phi}
\newcommand{\tolobjective}{\Psi}
\newcommand{\psize}{\texttt{m}}
\newcommand{\pset}{\texttt{P}}
\newcommand{\support}{\operatorname{support}}
\newcommand{\thres}{\texttt{$\tau$}}
\newcommand{\tol}{\eta}
\newcommand{\good}{\emph{$\tol$-good}\xspace}

\newtheorem{theorem}{Theorem}[section]
\newtheorem{corollary}[theorem]{Corollary}
\newtheorem{lemma}[theorem]{Lemma}
\newtheorem{observation}[theorem]{Observation}
\newtheorem{definition}[theorem]{Definition}
\algnewcommand{\LineComment}[1]{\State \(\triangleright\) #1}

\newcommand{\bigO}[1]{\mathcal{O}\left( #1 \right)}

\newcommand{\minB}{\min \{ \predError/n, f\} }

\newcommand{\GradedConsensus}{\textsc{graded-consensus}}

\newif\ifanonymous

\title{A Simple Phase-King Protocol with Predictions}
\ifanonymous
\author{}
\else
\author{
Antonia Giannoulea\textsuperscript{1},
George Papadoulis\textsuperscript{1,2},\\
Nikos Leonardos\textsuperscript{1},
Aris Pagourtzis\textsuperscript{1,2}\\[0.5em]
\small\textsuperscript{1}National Technical University of Athens\\
\small\textsuperscript{2}Archimedes/Athena RC\\[0.5em]
\small\texttt{antoniagian5@gmail.com},
\texttt{geopapadoulis@gmail.com},\\
\small\texttt{nikos.leonardos@gmail.com},
\texttt{pagour@cs.ntua.gr}
}
\fi
\date{}

\begin{document}

\maketitle

\begin{abstract}
Inspired by recent work (Ben-David, Dzulfikar, Ellen, and Gilbert, \textit{Byzantine Agreement with Predictions}, PODC 2025)
we study synchronous binary Byzantine Agreement in the learning-augmented setting, where each process receives
predictions about which processes are honest or Byzantine.
We present a deterministic Phase King protocol for the anauthenticated setting, 
combining early stopping with
a specifically designed prediction-aware king  selection procedure.
For at most $t<n/3$ Byzantine processes, the protocol handles
arbitrarily many prediction errors, improving upon the $\bigO{n^{3/2}}$ bound of Ben-David et al., 
and terminates within $\bigO{P}$
communication rounds, where
$P=1+\min\{\predError/n,f\}$,
$\predError$ is the total number of incorrect prediction
bits held by honest processes, and $f$ is the actual number
of Byzantine processes.
Our protocol uses $\bigO{n^2P}$ messages and
$\bigO{n^2P\log n}$ bits, avoiding the initial cubic bit cost
of exchanging complete prediction vectors.

In comparison to very recent independent work (Dufay,  Dzulfikar, and  Gilbert,  \textit{Predictions Can Only Help!
Communication Efficient Byzantine Agreement with Predictions}, DISC 2026),  
which achieves the same error tolerance and round complexity but better communication bounds, our construction provides a much simpler Phase King approach that completely avoids involved preprocessing constructions, rendering it more practical, especially for achieving agreement in time-evolving networks. We believe that the novel prediction-aware king selection technique of our scheme is of independent interest and can find further applications in learning-augmented distributed consensus.

\medskip
\noindent\textbf{Keywords:} Byzantine agreement $\cdot$ Algorithms with
predictions $\cdot$ Phase King $\cdot$ Round complexity $\cdot$
Communication complexity
\end{abstract}
\thispagestyle{empty}

\newpage
\setcounter{page}{1}

\section{Introduction}

Byzantine Agreement is a fundamental problem in distributed computing. A set
of processes must agree on a common value, even when some processes behave
arbitrarily.  The problem was introduced by Pease, Shostak, and
Lamport~\cite{pease1980agreement}. Any deterministic protocol requires $t+1$
rounds in the worst case, and early-stopping protocols, whose round complexity
depends on the actual number $f$ of Byzantine processes, require $\min\{f+2,
t+1\}$ rounds~\cite{dolev1990early}.  Classical protocols provide correctness without
knowing which processes are Byzantine. In practice, however, processes may have
additional information about the behavior of other participants. This raises a
natural question: Can such information help processes reach agreement faster?

This information is introduced as predictions. The idea behind prediction-based algorithms is the use of an external tool, likely based on AI or machine learning technique, to provide some predictive information. These data are based on historical or other external data independently of the algorithm in question.
In our setting, these are insights into the behavior of participating processes and predict those that may be faulty. Being able to correctly identify faulty processes in the protocol would enable faster agreement and termination.
However, predictions may have errors.
Therefore, when designing such algorithms, we must guarantee that even when predictions are inaccurate, the protocol should still be able to reach agreement with the same asymptotic round complexity as a protocol that does not rely on predictions.

The notion of predictions being used for solving Byzantine Agreement was
introduced by Ben-David, Dzulfikar, Ellen, and Gilbert in 
\cite{ben2025byzantine}. In their model, each process receives a prediction
vector indicating which processes are expected to be honest or Byzantine. These
vectors may differ between processes and may contain errors.  Their results show
that, while predictions do not improve communication complexity, which remains
bounded below by $\Omega(n+t^2)$ messages, they can significantly reduce round
complexity to $\bigO{1+\min\{\predError/n,f\}}$, where $\predError$ denotes the
total number of incorrect prediction bits held by honest processes.  
This means that accurate predictions improve the round complexity, while inaccurate ones do not worsen it beyond the $\bigO{f+1}$ rounds that can be achieved without predictions. They also show a lower bound of $\Omega(1+\min\{\predError/n,f\})$ rounds for deterministic protocols with classification predictions, making their result tight.

In their unauthenticated construction, this bound is guaranteed only for
$\predError=\bigO{n^{3/2}}$.  Extending this guarantee to a quadratic number of prediction errors was left as an open question.

\bigbreak \noindent \textbf{Our contribution.}
We give a specifically designed  Phase King protocol that solves
synchronous binary Byzantine Agreement with predictions with at most $t<n/3$
Byzantine processes,  terminates within $\bigO{1+\min\{\predError/n,f\}}$ communication rounds, and works for any number of prediction errors, i.e. for any $\predError\leq n^2$.

Our construction combines the majority-based classification of
Ben-David et al.~\cite{ben2025byzantine}, which yields a total of
$k=\bigO{\predError/n}$ misclassified processes, with a new king-selection
procedure. This procedure takes advantage of the classification
to select a common honest king within the first $k+1$ executions.
We integrate this procedure into Phase King by including both a
predetermined king and a prediction-based king in each iteration,
with graded consensus supporting agreement and early termination.

These two king steps result in termination in
$\bigO{P}$ communication rounds, where $P=1+\min\{\predError/n,f\}$.
Honest processes send $\bigO{n^2P}$ messages and
$\bigO{n^3+n^2P\log n}$ bits in total.

We further reduce communication by exchanging only a subset of each process's prediction vector, whose size depends on the prediction errors, rather than the complete vector. Since processes do not know the exact value of $\predError$, this approach is based on the guess-and-double strategy of \cite{ben2025byzantine}. 
The resulting protocol retains the $\bigO{P}$ round and $\bigO{n^2P}$ message bounds while
using $\bigO{n^2P\log n}$ bits, eliminating the initial cost of $\bigO{n^3}$ bits of full classification. The protocol and its analysis are presented in Appendix~\ref{app:smaller-communication}.

In independent and concurrent work, Dufay, Dzulfikar, and
Gilbert~\cite{dzulfikar2026communication} achieve stronger
communication bounds using extractor-based groupings derived
from expander graphs. Our protocol achieves the same 
round and error-tolerance bounds through a
specifically designed modification of Phase King,
without using  expander graphs or any other involved constructions. Therefore, despite 
its higher communication cost, it shows that prediction-aware 
agreement is possible without relying on heavy preprocessing thus providing a practical solution, especially meaningful in cases of time-evolving networks  where a repeated use of the protocol with varying participant sets may occur.
\section{Related Work}

The Classical Byzantine Agreement requires $n>3t$ in the unautheticated synchronous setting (\cite{pease1980agreement,lamport2019byzantine, berman1992bit}).
Deterministic protocols require $t+1$ rounds and $\Omega(nt)$ messages in the worst case (\cite{fischer1982lower, dolev1985bounds}).
One way to bypass these lower bounds is through early-stopping protocols, which adapt their termination time to the actual number of Byzantine faults $f$. These protocols achieve termination in $\min\{f+2,t+1\}$ rounds and $\bigO{n^2}$ bit complexity 
\cite{dolev1990early,berman1992early,abraham2015early, momose2021optimal, constantinescu2026adaptive, efron2026broadcast}.

Our work also relates to algorithms with predictions, which leverage imperfect information to improve performance while maintaining worst-case guarantees~\cite{mitzenmacher2020predictions}. Applications range from online algorithms~\cite{purohit2018improving,li2021load,azar2022online} to distributed graph problems~\cite{boyar2025distributed} and contention resolution~\cite{gilbert2021contention}. Unlike our work, these studies do not consider Byzantine failures.

Another related line of research concerns failure detectors, introduced by
Chandra and Toueg~\cite{chandra1996unreliable} for crash failures and extended
to Byzantine failures by Malkhi and Reiter~\cite{malkhi1997unreliable}.
Failure detectors provide processes with information about potentially faulty participants during protocol execution, allowing them to suspect or eventually identify processes exhibiting faulty behavior (\cite{doudou1998muteness, kihlstrom2003byzantine}.

Ben-David et al.~\cite{ben2025byzantine} introduced synchronous Byzantine
agreement with classification predictions in authenticated and
unauthenticated settings. Each process receives an additional binary vector
that includes predictions for which processes are honest or Byzantine. Processes exchange these predictions and reconcile them into classification vectors with a bounded number of misclassified processes. 
Let $\mathbf{B}$ denote the total number of incorrect prediction bits held by honest processes. Their protocols achieve
asymptotically optimal $\bigO{\min\{(\mathbf{B}/n)+1,f\}}$ round complexity 
when $\predError = \bigO{n^{3/2}}$.
They also prove an $\Omega(n+t^2)$ message lower bound for deterministic Byzantine agreement with predictions, even in executions where all predictions are correct. In comparison to their work we achieve the same round complexity while tolerating
any number of prediction errors, i.e. our protocol works for any  $\mathbf{B}\leq n^2$; however we deal only with the unauthenticated case. 

Dallot et al.~\cite{dallot2026resilient} investigate a complementary role for
predictions in Byzantine agreement: improving fault resilience rather than
reducing execution cost. They characterize trade-offs between the number of
faults tolerated when predictions are correct and the guarantees retained when
predictions are arbitrarily inaccurate, in both authenticated and
unauthenticated settings. They also study how resilience degrades as
prediction errors increase. Our work instead maintains resilience $t<n/3$ for
arbitrary predictions and uses prediction accuracy to improve round and
communication complexity.

In independent and concurrent work, Dufay, Dzulfikar, and
Gilbert~\cite{dzulfikar2026communication} developed protocols that achieve
optimal $\bigO{P}$ round complexity for arbitrary prediction errors, where
$P=1+\min\{\mathbf{B}/n,f\}$. Their unauthenticated construction combines
extractor-based grouping with group leader election to form committees sharing a
sufficiently large honest core. This achieves optimal resilience $t<n/3$ and
$\bigO{n^2\log n\log P}$ bit complexity without exchanging complete
prediction vectors.
Their extractor, taken from~\cite{dufay2026general}, is the incidence graph
of a regular Ramanujan graph, for which they rely on the deterministic
polynomial-time construction of Cohen~\cite{cohen2016ramanujan}. 
Instantiating their protocol therefore involves constructing this graph,
mapping processes to overlapping groups, and integrating these groups into the
agreement protocol. 
Although this construction is of theoretical interest, its practicality is limited, especially in cases of time-evolving networks that would require repeated use of the procedure. 
Our approach retains the familiar Phase King structure, using broadcasts,
prefix exchanges, and local updates to ordered lists, achieving the
same error tolerance and round complexity bounds, while
avoiding any graph construction or other involved preprocessing.
Our contribution thus offers a simpler route to prediction-aware agreement,
trading the higher by $\bigO{P/\log P}$
bit complexity for a fast king selection procedure that directly exploits advice from predictions.
This work is thus more applicable to practical settings, especially in dynamic situations where the set of players may vary from epoch to epoch.
\section{Preliminaries}

Table~\ref{tab:notation} (see Appendix~\ref{app:notation}) summarizes the notation which will be used throughout the paper. However each symbol is also introduced when first used.

\subsection{Model}

We consider a protocol in the permissioned setting consisting of $n$
processes $p_1, \dots, p_n$, identified by their indices; i.e., $i$ is the ID of process $p_i$.
In the rest of the paper we will use $i$ and $p_i$ interchangeably.
Up to $t<n/3$ processes may be Byzantine. We denote by $f
\leq t$ the actual number of Byzantine processes, $\Honest$ the set of honest
processes and $\Faulty$ the set of the Byzantine processes.

We consider a synchronous network. Every process can send a message to every
other process directly, and messages between honest processes arrive within the communication round.
The receiver can identify the sender through the incoming channel.
We use $\Broadcast(\msg)$ to denote sending $\msg$ to all processes
and $\Receive()$ to denote the messages received during the current round.
We count point-to-point messages; therefore, one broadcast by one process
consists of $\bigO{n}$ messages.

We consider binary Byzantine Agreement, with input and output
domain $\inputdomain$. Each process $p_i$ starts with an input
$x_i\in\inputdomain$.

Each process $p_i$ also receives a prediction vector $\preds_i \in \{0,1\}^n$ as input,  where
$a_i[j] = 0$ denotes a prediction that process $p_j$ is Byzantine and $a_i[j] =
1$ denotes a prediction that process $p_j$ is honest. We underline that each process
may receive different predictions.

We denote by $\predError$ the total number of incorrect prediction bits held by
honest processes. Let $\type(p_j) \in \{0,1\}$ denote the actual type of process
$p_j$, where 0 means Byzantine and 1 honest. Then $$\predError = \sum_{p_i \in
\Honest} \sum_{j=1}^{n} \mathbbm{1}[a_i[j] \neq \type(p_j)]$$


\subsection{Byzantine Agreement}
The problem of Byzantine Agreement is defined as follows: each process $p_i$ has input a value $x_i$ and 
produces an output $y_i$ satisfying:
\begin{itemize}
    \item \emph{Validity}: If every honest process has the same input $v$, then every honest process outputs $v$.
    \item \emph{Agreement}: All honest processes output the same value.
    \item \emph{Termination}: Eventually every honest process terminates and outputs a value.
\end{itemize}

The round complexity of a Byzantine Agreement protocol is the number of rounds it takes for all honest processes to terminate,
while the message/bit complexity is the total number of messages/bits sent by honest processes during the execution of the 
protocol. 

We will also use a weaker version of the problem called \textsc{graded-consensus}. In this version,
each process $p_i$ has input a value $x_i$ and produces an output $(y_i, g_i) \in \{0,1\}\times\{0,1\}$ satisfying \emph{Termination} and:
\begin{itemize}
    \item \emph{Validity}: If every honest process has the same input $v$, then every honest process outputs $(v,1)$.
    \item \emph{Graded Agreement}: If an honest process outputs a value $v$ with grade $1$, i.e. $(v,1)$, then all honest
        processes should output $(v, \ast)$.
\end{itemize}

We instantiate \textsc{graded-consensus} with the binary weak
validator of \cite{lenzen2022recursive}
, interpreting its output $\mathit{strong}_i$
as grade $g_i$. For $t<n/3$, \cite{lenzen2022recursive} establish validity and graded agreement and each invocation
terminates after two communication rounds, during which each
honest process sends at most $2n$ bits.
\section{Classification and King Selection}

This section presents two components of our Byzantine Agreement
protocol (\cref{alg:ba} \cref{sec:main}). Classification
(\cref{alg:classify}) aggregates prediction vectors using majority
votes. King-selection (\cref{alg:propose}) uses
the resulting classifications to exchange and update proposals
in search of a common honest king.

\subsection{Classification Algorithm}

The initial prediction vectors of honest processes may differ.
Algorithm~\ref{alg:classify} reconciles these predictions by having every
process broadcasts its prediction vector and computes each classification bit
using a majority vote. This procedure limits the number of inconsistent
classifications among honest processes, as formalized by the lemma below.

\begin{algorithm}[H]
\caption{Classification Algorithm (code for process $p_i$)}
\label{alg:classify}
\begin{algorithmic}[1]
    \Procedure{Classify}{$\preds_i$} \Comment{prediction vector $\preds_i \in \{0,1\}^n$} 
        \State $\Broadcast(\preds_i$)
        \State $R_i \gets \Receive()$ \Comment{$R_i$: set of prediction vectors}

        \For{$j = 1$ to $n$}
            \State $\mathrm{honesty}_j \gets \left|\left\{a \in R_i : a[j]=1\right\}\right|$
            \Comment{\#predictions classifying $p_j$ as honest}
            \If{$\mathrm{honesty}_j \ge \left\lceil\frac{n+1}{2}\right\rceil$}
                $c_i[j] \gets 1$
            \Else{
                $c_i[j] \gets 0$
            }
            \EndIf
        \EndFor

        \State \Return $\clas_i$ \Comment{classification vector $\clas_i \in \{0,1\}^n$}
    \EndProcedure
\end{algorithmic}
\end{algorithm}

\begin{definition}[Misclassified Process]

    A process $p_j$ is \emph{misclassified} if there exists an honest process
    $p_i \in \Honest$ such that $c_i[j] \neq \type(p_j)$.

\end{definition}

We denote by $k$ the total number of misclassified processes.

\begin{lemma}[\cite{ben2025byzantine}] \label[lemma]{lem:k-bound}

    If $f < \epsilon n$ for some constant $0 < \epsilon < 1/2$, then $k = \bigO{\predError/n}$.

\end{lemma}

\begin{definition}[Common Process]\label[definition]{def:common-process}

    A process $p_j$ is \emph{common} if all honest processes assign the same
    classification to $p_j$, i.e.,
    $\forall p_i,p_{i'} \in \Honest,
        c_i[j] = c_{i'}[j].
    $
    A process $p_j \in \Honest$ is a \emph{common honest process} if every honest process classifies it as honest .
\end{definition}

For each process $p_i$, let $\Honest_i$ be the set of
ids that $p_i$ classifies as honest
after executing \cref{alg:classify}.
Formally,
$
    \Honest_i = \{j \in \{1,\dots,n\} \mid c_i[j] = 1\}.
$
\subsection{Classification-based King Selection}\label{sec:king-selection}

Algorithm~\ref{alg:propose} describes the king-selection procedure executed by
each process at every Phase King round. 
Its goal is to ensure that, in some round, all honest processes select the same honest king. We start by an informal discussion of the most important properties of the algorithm and we will then proceed to its formal analysis in Subsection~\ref{sec:king_analysis}.

Honest processes search for such a king by considering ids they have classified as honest in increasing order. 
Each process initially proposes the least id in its list of honest ids and once it has proposed an id in a certain round, it never decreases its  proposal in consequent rounds. 
This monotonicity allows a process to discard candidates once it learns that another honest process has proposed  a larger id, since those candidates can no longer be proposed by all honest processes.
The first rule (line~\ref{line:pos_change}) uses a threshold of $t+1$ votes to obtain this evidence. A process counts all votes for ids in its candidate list, and selects the largest id $j$ such that the sum of votes for candidates with an id  larger or equal to $j$ is at least $t+1$. 
Since at most $t$ votes come from faulty processes, this threshold guarantees that at least one honest process has already proposed a candidate with an id larger or equal to $j$. Therefore, moving to  candidate with id $j$  safely skips smaller ids.

However, this rule alone does not guarantee progress. All honest processes could remain at the same proposal indefinitely. The second rule (line~\ref{line:pos_change_strictly}) breaks such a stalemate. A process advances its next proposal beyond its selected king whenever it has cumulatively received at least $2t+1$ votes for that king or for processes with larger ids, counting for this step votes for all processes regardless of their classification.
This ensures that the minimum honest proposal will strictly increase, since every process whose proposed id is the minimum among honest processes receives at least $n-t\geq 2t+1$ votes for this or larger ids.

\begin{figure}[H]
    \centering
\begin{tikzpicture}[x=0.72cm,y=0.42cm,
    hon/.style={circle,fill=blue!70!black,inner sep=0pt,minimum size=5pt},
    byz/.style={cross out,draw=red!75!black,very thick,inner sep=0pt,minimum size=5pt}]

    \fill[red!7] (6.5,0) rectangle (10.7,10.8);

    \foreach \x in {1,...,10}
        \draw[gray!25] (\x,0) -- (\x,10.6);

    \draw[->] (0,0) -- (11.2,0) node[right] {\small proposed id};
    \draw[->] (0,0) -- (0,11.4) node[above] {\small sender};

    \foreach \x/\l in {1/1, 2/2, 4/\cdots, 6/\common, 10/n}
        \draw (\x,0.15) -- (\x,-0.15) node[below] {$\l$};
    \foreach \y/\l in {1/p_1, 2/p_2, 10/p_n}
        \draw (0.08,\y) -- (-0.08,\y) node[left] {$\l$};
    \node[left] at (0,5.8) {$\vdots$};

    \draw[dashed,thick] (6.5,0) -- (6.5,10.8);
    \draw[gray!70!black,thick] (6,5.5) ellipse (0.42 and 5.25);

    \foreach \x/\y in {6/1, 3/2, 6/4, 6/5, 2/6, 6/8, 5/10}
        \node[hon] at (\x,\y) {};
    \foreach \x/\y in {8/3, 6/7, 10/9}
        \node[byz] at (\x,\y) {};

    \draw[decorate,decoration={brace,amplitude=5pt}]
        (0.6,11.3) -- (6.4,11.3)
        node[midway,above=5pt,align=center] {\small all honest proposals \\ \small (proposed ids $\leq \common$)};
    \node[align=center] at (8.6,12.6) {\small only Byzantine \\ \small proposals\\[-2pt] \small (at most $t$ votes)};
    \node[align=center,fill=white,inner sep=1pt] (lbl) at (3.3,8.2)
        {\small $\geq t+1$ honest\\[-2pt] \small votes for $\common$};
    \draw[gray!70!black,->] (lbl.east) -- (5.55,7.6);
    \draw[decorate,decoration={brace,mirror,amplitude=5pt}]
        (5.6,-1.4) -- (10.4,-1.4)
        node[midway,below=5pt] {\small $\geq 2t+1$ votes  for ids $\geq \common$};

    \node[hon] at (11.9,8) {};
    \node[right] at (12.1,8) {\small honest vote};
    \node[byz] at (11.9,6.8) {};
    \node[right] at (12.1,6.8) {\small Byzantine vote};
    \node[right,align=left] at (11.6,3.6) {\small Round $r^\ast$:\\
        \small every honest $p_j$\\ \small selects $\common$};
\end{tikzpicture}
    \caption{
        Illustration of round $r^*$ in which an honest process advances beyond $\common$: there are no honestly proposed ids larger than $\common$, therefore at least $t+1$ honest processes have proposed $\common$ in this round.
    }
    \label{fig:race}
\end{figure}
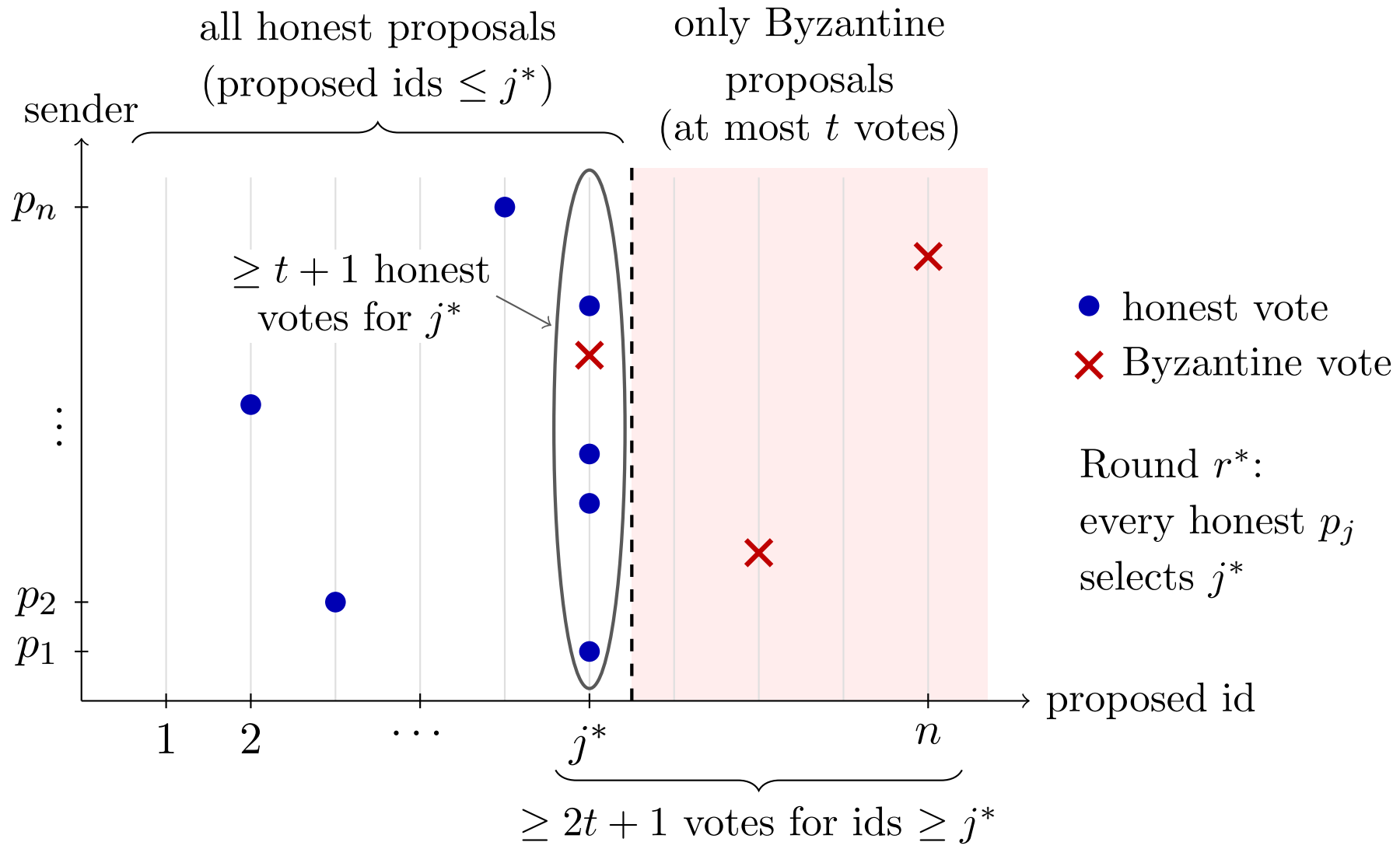

Since progress is guaranteed, it remains to show that no honest process advances beyond the common honest id $\common$ unless all honest processes select it as king in that same round. Consider the first round in which an honest process advances its next proposal beyond $\common$. In this round, no honest process yet proposes an id larger than $\common$, so the $2t+1$ threshold must include votes from at least $t+1$ honest processes proposing $\common$ itself (see~\cref{fig:race} for an illustration). These votes ensure that $\common$ satisfies the first rule in the computation  of every honest process. Moreover, no larger candidate can satisfy that rule, since only faulty processes can propose larger ids. Thus, all honest processes select $\common$ as king in that round.

\begin{algorithm} [H] 
\caption{King Proposal and Selection (code for process $p_i$)}
\label{alg:propose}
\begin{algorithmic}[1]
    \State \textbf{Input:} ordered leader vector $\lead_i$, current position $\pos_i \in \{1,\dots, |\lead_i|\}$
    \State \textbf{Output:} selected king ID $\king \in \{1,\dots,n\}$, updated position $\pos_i^\prime$
    \Procedure{Propose}{$\lead_i, \pos_i$}
        \State $\Broadcast(\lead_i[\pos_i])$

        \State $\texttt{votes} \gets \Receive()$
        \Comment{$\texttt{votes}[u]$ = \#votes for process $u$}
        \State $cnt \gets 0, \king  \gets \lead_i[\pos_i], \pos_i^\prime \gets \pos_i$
        \label{line:start_king}
        \For{$j \gets n$ \textbf{down to} $\king$}
            \If{$j\in \lead_i$}
                \State $cnt \gets cnt + \texttt{votes}[j]$
            \EndIf
            \If{$cnt \geq t+1$} \label{line:king_condition}
                \State $\king \gets j$
                \Comment{$j\in \lead_i$}
                \label{line:king_select}
                 \State $\pos_i^\prime \gets \ell \text{ such that } \lead_i[\ell] = \king$
            \label{line:pos_change}
                    \State \textbf{break}
            \EndIf
        \EndFor
        \If{ $\texttt{sum}(\texttt{votes}[\king..n ]) \geq 2t + 1 \land \pos_i^\prime < |\lead_i|$} \label{line:proceed_condition}
            \State $\pos_i^\prime \gets \pos^\prime_i+1$
            \Comment{$\min\{\ell : \lead_i[\ell] > \king\}$}
            \label{line:pos_change_strictly}
        \EndIf 

        \State \Return $(\king, \pos_i^\prime)$ \label{line:return}
    \EndProcedure
\end{algorithmic}
\end{algorithm}

\subsection{Analysis}\label{sec:king_analysis}

Let $n$ be the number of processes, up to $t < n/3$ of which may be Byzantine.
Let $\Misclas$ be the set of ids misclassified by at least one honest
process. Thus $|\Misclas| = k$.
Throughout this analysis, every honest process executes
\textsc{Propose} once in every round, using its fixed list
$\lead_i$. For every $r\geq 1$, its invocation in round $r+1$
uses the updated position returned by its invocation in round $r$.

Each process counts at most one valid proposed id per sender, including
its own proposal.  Honest proposals are counted by every honest process and at
most $t$ counted votes at any recipient are Byzantine.  The loop that selects a
king scans ids downwards from $n$ to the process's current proposal.

Let $q_i(r) = \lead_i[\pos_i(r)]$ be the proposal of $p_i$ in round $r$ (i.e the value of $\king$ in Line~\ref{line:start_king} of \cref{alg:propose}), and $V_i^r(j)$ be
the number of votes for id $j$ received by $p_i$ in round $r$ (i.e. the value $\texttt{votes}$ in \cref{alg:propose}).
Define the \emph{suffix count} $U_i^r(s)$ and the
\emph{candidate suffix count} $\widehat U_i^r(s)$ by
\begin{align*}
    U_i^r(s)=\sum_{j=s}^{n}V_i^r(j),
    &&
    \widehat U_i^r(s)=\sum_{\substack{j=s\\ j\in\lead_i}}^{n}V_i^r(j).
\end{align*}

Note that the suffix count corresponds to $\texttt{sum}(\texttt{votes}[\king..n ])$,
while the candidate suffix count corresponds to the value of $cnt$ in \cref{alg:propose}.

Let $s_i(r)$ be the king selected by $p_i$ during
round $r$, i.e. coincides with the value of $\king$ in Line~\ref{line:return} of \cref{alg:propose}.  It is the largest id not smaller than $q_i(r)$ whose
candidate 
suffix count is at least $t+1$. If no such id
exists, $s_i(r)=q_i(r)$.

More formally, let $$A_i(r) = \{s \in \{q_i(r), \dots, n\} : \widehat U_i^r(s) \geq t+1\}.$$
Then 
$$ s_i(r) = \begin{cases}
    \max A_i(r) , & \text{if } A_i(r)\neq\varnothing, \text{(Line~\ref{line:king_select} of \cref{alg:propose})}\\
    
    q_i(r), & \text{otherwise (Line~\ref{line:start_king} of \cref{alg:propose})}.
    \end{cases}
$$



The next proposal is determined by the updated position of $\lead_i$.
The first case of the following update rule is called the
\emph{advancing branch}:
\[
    \pos_i(r+1) = \begin{cases}
        \pos_i(r) + 1, & \text{if } U_i^r(s_i(r)) \geq 2t+1\\
        & \quad\text{and } s_i(r) < \max(\lead_i),\\
        \text{the unique position of } s_i(r) \text{ in } \lead_i, & \text{otherwise}.
    \end{cases}
\]



Now we are ready to prove the following lemmas.




\begin{lemma} \label[lemma]{lem:common_existance}
    If $k<n-t$, then there exists at least one common honest process.
\end{lemma}

\begin{proof}

    There are at least $n-t$ honest processes. If none were common honest, then
    every honest process would be misclassified by at least one honest process
    and therefore belong to $\Misclas$. This implies $|\Misclas| \geq n-t > k$,
    contradicting $|\Misclas| = k$.

\end{proof}

In all the following lemmas, we consider the case where $k<n-t$ (if not, agreement will be reached by the predetermined kings). 
Let $\common$ denote the smallest id of a common honest process.


\begin{lemma}[Density]\label[lemma]{lem:density}
    There are at most $k$ ids smaller than $\common$ that are classified
    as honest by at least one honest process. More formally, let $\densityset$ be the set of these ids
    $$ \densityset = 
        \left\{j<\common : j\in\bigcup_{p_i\in\Honest}\Honest_i\right\}.
    $$
    Then $|\densityset| \leq k.$
\end{lemma}

\begin{proof}

    Let $j \in \Honest_i$ for some honest $p_i$. If $j \notin \Misclas$,
    then $p_j \in \Honest$ and every honest process classifies it as honest.
    Thus, $j$ is a common honest id.
    Therefore, each element of $\densityset$ is common honest or
    misclassified. By the minimality of $\common$, $\densityset$ contains no
    common honest process, since all its elements are smaller than $\common$.
    Therefore $\densityset \subseteq \Misclas$ and 
    $|\densityset| \leq |\Misclas| = k$.

\end{proof}

\begin{lemma}[Monotonicity]
\label{obv:monotonicity}
    For every honest $p_i$ and every round $r$,
    $$ q_i(r+1) \geq s_i(r) \geq q_i(r). $$
\end{lemma}

\begin{proof}
    
    Selection either keeps $q_i(r)$ or chooses a larger id, so $s_i(r) \geq
    q_i(r)$.  By definition $q_i(r+1) = \lead_i[\pos_i(r+1)] $. The updated
    position either stays at the selected king or advances to a later entry of
    the increasing list $\lead_i$, so $q_i(r+1) \geq s_i(r)$.

\end{proof}

\begin{lemma}[Progress]\label[lemma]{lem:progress}

    Let $m_r = \min_{p_i \in \Honest} q_i(r)$ be the minimum honest proposed
    id in round $r$. If $m_r < \common$, then $m_{r+1} > m_r$.

\end{lemma}

\begin{proof}

    Fix a round $r$ with $m_r < \common$, and consider any honest process $p_i$.
    If $q_i(r) > m_r$, \cref{obv:monotonicity} gives $q_i(r+1) \geq q_i(r) > m_r$.
    Otherwise, $q_i(r) = m_r$. By \cref{obv:monotonicity} $s_i(r) \geq m_r$. 
    
    If $s_i(r) > m_r$, then $ q_i(r+1) \geq s_i(r) > m_r$.
    Finally, suppose $s_i(r) = m_r$. Since $m_r$ is the minimum honest proposed 
    id, then every honest process proposes an id $\texttt{ID} \geq m_r$.
    Hence $U_i^r(m_r) \geq n-t$ and $n-t \geq 2t+1$ since $n>3t$.
    Both conditions of the advancing branch hold:
    $U_i^r(m_r)\geq 2t+1$ and
    $s_i(r)=m_r<\common\leq\max(\lead_i)$.
    Hence $\pos_i(r+1)=\pos(r)+1$.
    It follows that $q_i(r+1)>m_r$.
    
    Therefore, $q_i(r+1) > m_r$ for every honest process $p_i$, and hence
    $ m_{r+1}=\min_{p_i\in\Honest}q_i(r+1)>m_r$.

\end{proof}

\begin{definition}[Crossing round]\label[definition]{def:crossing}
A round $r$ is a \emph{crossing round} if
$q_i(r+1)>\common$ for some honest process $p_i$.
The \emph{first crossing round} is the smallest such round,
if one exists.
\end{definition}

\begin{corollary}\label[corollary]{cor:crossing}
Either a crossing round exists and the first crossing round
is at most $k+1$, or there is a round $r\leq k+1$ in which
every honest process proposes $\common$:
$$
    q_i(r)=\common
    \qquad \text{for every honest }p_i.
$$
\end{corollary}

\begin{proof}
    If the first crossing round is at most $k+1$,
    the first alternative holds.
    Otherwise, the proposed id of each honest proposal is initially at most $\common$. Since no crossing occurs
    within the first $k+1$ rounds, every honest proposal remains at most
    $\common$ throughout these rounds.
    Whenever $m_r<\common$, there is an honest process $p_i$
    with $m_r=q_i(r)\in\Honest_i$.
    Thus $m_r\in\densityset$.
    By \cref{lem:density}, $m_r$ can therefore take
    at most $k$ distinct values below $\common$.
    By \cref{lem:progress}, it strictly increases whenever it
    is below $\common$. Therefore, it must equal $\common$
    in some round $r\leq k+1$.
    In that round, every honest proposal is at least the
    minimum $\common$ and at most $\common$. Hence
    $q_i(r)=\common$ for every honest $p_i$.
\end{proof}

\begin{lemma}[Convergence]\label[lemma]{lem:convergence}
If a crossing round exists, let $r^\ast$ be the
first crossing round. Then every honest process selects
$\common$ as king in round $r^\ast$:
\[
    s_i(r^\ast)=\common
    \qquad\text{for every honest }p_i.
\]
\end{lemma}
\begin{proof}
    By the definition of a crossing round, there exists an honest
    process $p_i$ with $q_i(r^\ast+1)>\common$.
    Since this is the first crossing round, then
    $ q_j(r^\ast)\leq\common \text{ for every honest }p_j.$
    Therefore, all votes for ids greater than $\common$ in round $r^*$
    originate from Byzantine processes. This means that
    $\forall p_j \in \Honest, s > \common: \ \widehat U_j^{r^*}(s) 
    \leq U_j^{r^*}(s) \leq t$, and thus no id $s>\common$ belongs to $A_j(r^\ast)$.
    If $A_j(r^\ast)\neq\varnothing$, then
    $s_j(r^\ast)=\max A_j(r^\ast)\leq\common$.
    If $A_j(r^\ast)=\varnothing$, then
    $s_j(r^\ast)=q_j(r^\ast)\leq\common$.
    Thus, $s_j(r^\ast)\leq\common$ for every honest $p_j$.

    Therefore, for the witnessing process $p_i$, the update rule must take
    the advancing branch. Otherwise, it would give
    $q_i(r^\ast+1)=s_i(r^\ast)\leq\common$,
    contradicting $q_i(r^\ast+1)>\common$.
    Consequently, the advancing condition holds, and in particular
    $U_i^{r^\ast}(s_i(r^\ast))\geq 2t+1.$

    If $s_i(r^\ast)<\common$, then, since $\common\in\lead_i$,
    the first entry of $\lead_i$ larger than $s_i(r^\ast)$
    is at most $\common$. The advancing branch would therefore give
    $q_i(r^\ast+1)\leq\common$, a contradiction.
    Since $s_i(r^\ast)\leq\common$ was established above,
    we conclude that $s_i(r^\ast)=\common$. Substituting into
    the threshold bound gives
    \[
        U_i^{r^\ast}(\common)\geq 2t+1.
    \]

    At most $t$ of these votes originate from Byzantine processes,
    so at least $t+1$ originate from honest processes.
    Since no honest process proposed an id greater than
    $\common$, all these honest votes are for $\common$ itself.
    Since $\common\in\Honest_j$,
    $$ \forall p_j \in \Honest: \quad \widehat U_j^{r^\ast}(\common)\geq t+1.$$

    Since $q_j(r^\ast)\leq\common$ and
    $\widehat U_j^{r^\ast}(\common)\geq t+1$,
    we have $\common\in A_j(r^\ast)$.
    Together with the earlier exclusion of larger ids,
    this implies $\max A_j(r^\ast)=\common$.
    Hence $s_j(r^\ast)=\common$ for every honest $p_j$.

\end{proof}

\begin{theorem}\label{theorem:king_agreement}
If $k<n-t$, then among the first $k+1$ executions of
\cref{alg:propose}, there is one in which all honest processes
select the same honest king.
\end{theorem}

\begin{proof}
    By \cref{lem:common_existance}, a common honest process $\common$ exists.
    By \cref{cor:crossing}, either the first crossing round
    is at most $k+1$, or all honest processes propose $\common$
    in some round $r\leq k+1$.

    In the first case, \cref{lem:convergence} ensures that all
    honest processes select $\common$ in the crossing round.
    In the second case, let $r$ be a round in which
    $q_i(r)=\common$ for every honest $p_i$.
    Every honest recipient receives at least $n-t$ votes for
    $\common$. Since $\common\in\Honest_i$, we have
    $\widehat U_i^r(\common)\geq n-t\geq t+1$.
    Hence $\common=q_i(r)\in A_i(r)$.

    All votes for ids greater than $\common$ originate
    from Byzantine processes. Therefore,
    $\widehat U_i^r(s)\leq t$ for every $s>\common$,
    so no such $s$ belongs to $A_i(r)$.
    Consequently, $\max A_i(r)=\common$, and thus
    $s_i(r)=\common$ for every honest $p_i$.

    Since $\common$ is honest, the theorem follows.
\end{proof}

\section{A protocol for Byzantine Agreement with Predictions} \label{sec:main}

The goal of the protocol is that every honest process decide on the same value and terminate. We will say that a process `records a decision' to express that the process has reached a decision that will never change afterwards, and we will say that a process `returns' to express that the process terminates. 

The protocol combines two ways of finding a common honest king: one guarantees progress regardless of prediction quality, while the other exploits accurate predictions to reach agreement sooner.
To this end, each Phase King round $r$ contains two king selection steps.
The first uses the predetermined process $p_r$ as king, which guarantees a common honest king within the first $f+1$ rounds regardless of prediction quality.
The second uses the king returned by \textsc{Propose} (\cref{alg:propose}), which guarantees a common honest king within the first $k+1$ rounds by \cref{theorem:king_agreement}.

Each king selection is followed by a graded-consensus execution. This allows to preserve agreement once it has been reached, without being affected by the next king selection step. Thus, the protocol benefits from whichever king-selection method provides a common honest king first.

To detect when it is safe to decide, each round ends with an additional graded-consensus execution. A process that obtains a value with grade $1$ records the returned value as its decision (Line~\ref{line:decision} of \cref{alg:ba}). It then participates in one more round, allowing the remaining honest processes to record the same decision, before terminating. As we show below, all honest processes have recorded the same decision by the time any honest process returns.



\begin{algorithm}[H]
\caption{Byzantine Agreement (code for process $p_i$)}
\label{alg:ba}
\begin{algorithmic}[1]
\Procedure{ba-with-predictions}{$x_i,\preds_i$}
    \State $\clas_i \gets \textsc{classify}(\preds_i)$, $\Honest_i \gets \{j : c_i[j] = 1\}$, $\lead_i \gets \textsc{sort}(\Honest_i)$
    \If{$\lead_i=\varnothing$} $\lead_i\gets[i]$
    \EndIf
    \State $\pos_i \gets 1, v_i \gets x_i, \mathit{decided_i} \gets \textbf{false}$

    \For{$r = 1$ to $t + 1$}
        \State $(v_i,g_i) \gets \textsc{graded-consensus}(v_i)$ \label{line:gc-1}
        \State $\Broadcast(v_i)$
        \If{$g_i = 0$}
            $v_i \gets v_r$ \Comment{value received from $p_r$ (king)} 
        \EndIf

        \State $(v_i,g_i) \gets \textsc{graded-consensus}(v_i)$ \label{line:gc-2}
        \State $(\king_i, \pos_i) \gets \textsc{Propose}(\lead_i, \pos_i)$
        \State $\Broadcast(v_i)$
        \If{$g_i = 0$}
            $v_i \gets v_{\king_i}$\Comment{value received from $p_\king$} 
        \EndIf

        \State $(v_i,g_i) \gets \textsc{graded-consensus}(v_i)$

        \If{$\mathit{decided_i}$}
            \Return $\mathit{decision_i}$
        \EndIf

        \If{$g_i = 1$}
            \State $\mathit{decision_i} \gets v_i$ \label{line:decision}
            \Comment{record decision $v_i$}
            \State $\mathit{decided_i} \gets \textbf{true}$
        \EndIf
    \EndFor

    \State \Return $\mathit{decision_i}$
\EndProcedure
\end{algorithmic}
\end{algorithm} 

In all the following lemmas we consider the case where $t<n/3$. In addition,
in the following analysis, a \emph{round} denotes one iteration
of the main loop of \cref{alg:ba}.
Each iteration contains three instances of graded-consensus,
two value broadcasts, and one invocation of \textsc{Propose}.
Since each subprotocol takes a constant number of synchronous
communication rounds, each iteration takes $\bigO{1}$ communication
rounds. Thus, bounds on loop iterations yield the same
asymptotic bounds on communication rounds.

\begin{lemma}[Preservation of a common value]\label[lemma]{lem:preservation-simple}
Suppose all honest processes participate in round $r$ and hold the same value $v$
before some graded-consensus instance of round $r$. Then that instance and every
later graded-consensus instance in round $r$ return $(v,1)$ at every honest
process, and all honest processes finish the round holding $v$.
\end{lemma}
\begin{proof}
    By the validity of graded consensus, in the first instance of graded-consensus 
    all honest processes return $(v,1)$. Thus no honest process adopts the
    predetermined king's value. So, in the second instance all honest processes
    start with the same input $v$ and again from validity return $(v,1)$.
    No honest process adopts the proposed king's value, so the final instance again 
    receives only $v$ from honest processes and returns $(v,1)$.
\end{proof}

\begin{lemma}[Validity]\label[lemma]{lem:validity-simple}
    If every honest process starts with input $v$, then every
    honest process returns $v$.
\end{lemma}

\begin{proof}
    By \cref{lem:preservation-simple} all honest processes in the final graded
    consensus in the first round return $(v,1)$. Thus they all record decision
    $v$.  If $t=0$, the loop ends and they all return $v$. Otherwise, all honest
    processes participate in the second round, preserve $v$ and since 
    $\mathit{decided}_i = \textbf{true}$ for every $p_i \in \Honest$ they return their stored decision $v$.
\end{proof}

\begin{definition}[Common honest king]
\label[definition]{def:common-honest-king}
A round $r$ has a \emph{common honest king} if either $p_r$
is honest, or every honest process's invocation of
\textsc{Propose} in that round returns the same honest id.
\end{definition}

\begin{lemma}[Common Honest King Agreement]\label[lemma]{lem:agreement-in-chk-simple}
    If a round has a common honest king, then all honest processes have the
    same value at the end of that round. In particular, there exists a value $v$
    such that the final graded-consensus instance returns $(v,1)$ at every honest process.
\end{lemma}
\begin{proof}
    Let $r$ be a round with a common honest king, and choose
    a king step in which all honest processes use the same
    honest king. Consider the graded-consensus instance
    immediately preceding that step: the first instance for
    the predetermined king (Line~\ref{line:gc-1}), or the second
    instance for the proposed king (Line~\ref{line:gc-2}).

    If some honest process returns $(v,1)$ for some value $v$
    from this instance, then the graded-agreement property
    ensures that every
    honest process returns $(v, \ast)$, including the honest king. Processes
    with grade 1 retain $v$, while those with grade 0 adopt the honest king's
    value, which is also $v$.
    Otherwise all honest processes have grade 0 and adopt the same value sent by 
    the honest king. Thus all honest processes hold one common value after the 
    king's value adoption step.


    By \cref{lem:preservation-simple}, every remaining graded-consensus
    instance of round $r$ returns $(v,1)$ at every honest process, and all
    honest processes finish the round having decided on $v$.
\end{proof}

\begin{lemma}[Decision Propagation]\label[lemma]{lem:decision}
    If an honest process records a decision $v$ at round $r$, then all honest 
    processes hold $v$ at the end of that round. Also, if $r<t+1$ then 
    by the end of round $r+1$ all honest processes have recorded $v$.
\end{lemma}
\begin{proof}
    A decision is recorded only after the final graded consensus returns grade 1.
    By the graded-agreement property if an honest process returns $(v,1)$ then 
    every honest process returns $(v, \ast)$. Therefore,
    if $r < t+1$ then all honest processes start round $r+1$ with the same value $v$, so by \cref{lem:preservation-simple} every honest process that has not yet recorded a decision records $v$ in round $r+1$.
    If $r = t+1$ then all honest processes end with value $v$.
\end{proof}

\begin{lemma}[Termination]\label[lemma]{lem:termination-simple}
Every honest decides and terminates by the end of round $R=\min\{t+1,\,f+2,\,k+2\}$.
\end{lemma}
\begin{proof}
    We establish two independent upper bounds on termination.
    \begin{itemize}
        \item \textbf{Predetermined selection:} 
        Let $h$ be the first iteration whose predetermined king $p_h$ is honest.
        Among the first $f+1$ ids at least one is honest, so $h \leq f+1
        \leq t+1$.

        If some honest process records a decision in a round $r<h$,
        then by \cref{lem:decision}, every honest process has
        recorded that decision by round $r+1$.
        Thus, all honest processes return by round
        $\min\{r+2,t+1\}\leq\min\{h+1,t+1\}$.

        Otherwise, no honest process has recorded a decision
        before round $h$, so all honest processes participate
        in that round.
        In round $h$,
        by \cref{lem:agreement-in-chk-simple} the final graded-consensus
        returns grade 1 in every honest process, so every honest process
        records a decision. 
        If $h\leq t$ then in round $h+1 \leq \min\{f+2,t+1\}$ every honest
        process terminates. 
        If $h=t+1$, every honest process returns after the loop.

        \item \textbf{Selection by proposals:}
        For the prediction bound, it suffices to consider $k<t$:
        if $k\geq t$, then $f\leq t\leq k$, so the
        predetermined-selection bound suffices.
        Since $k<t<n-t$, the classification requirement of
        \cref{theorem:king_agreement} is satisfied.
        There is therefore a common honest process whose id
        appears in every honest process's leader list.

        If some honest process records a decision in a round
        $r\leq k$, then by \cref{lem:decision}, every honest
        process has recorded that decision by round $r+1$.
        The algorithm therefore makes every honest process
        return by round $r+2\leq k+2$.

        Otherwise, no honest process returns before round $k+2$.
        Thus, all honest processes participate in the first
        $k+1$ invocations of \textsc{Propose}, and
        \cref{theorem:king_agreement} guarantees a common honest
        king within these rounds.
        By \cref{lem:agreement-in-chk-simple}, every honest
        process records a decision by round $k+1$ and returns
        by round $k+2\leq t+1$.
    \end{itemize}

    Combining the two bounds, every honest process records a decision and returns it by the end of round $R = \min\{t+1,f+2,k+2\}$.
\end{proof}

\begin{lemma}[Agreement]\label[lemma]{lem:agreement-simple}
    All honest processes return the same value.
\end{lemma}
\begin{proof}
    Let $T = t+1$, and let $r$ be the earliest round in which an honest process records a decision.  
    Such a round exists by \cref{lem:termination-simple}. Let $v$ be a decision recorded by an honest process in
    round $r$.

    Suppose $r<T$. 
    By \cref{lem:decision}, every honest process holds $v$
    at the end of round $r$, and every honest process has
    recorded $v$ by the end of round $r+1$.
    In round $r+1$, the processes that recorded $v$ in round $r$ return $v$,
    and every other honest process records $v$.
    If $r+1 = T$, the remaining processes return $v$ after the loop.
    Otherwise, they return their recorded decision $v$ at the end of
    round $r+2$, regardless of the outcome of that round.
    So every honest process returns $v$.

    It remains to consider the case $r=T$.
    No honest process has recorded a decision earlier,
    so all honest processes participate through round $T$.
    Suppose $p_j$ were honest for some $1\leq j\leq t$.
    Then round $j<T$ would have a common honest predetermined king.
    By \cref{lem:agreement-in-chk-simple}, every honest process
    would record a decision in round $j$, contradicting the
    choice of $r$.
    Therefore, $p_1,\dots,p_t$ are all Byzantine.
    Since $f\leq t$, the predetermined king $p_T=p_{t+1}$ is honest.
    By \cref{lem:agreement-in-chk-simple}, the final graded-consensus
    instance of round $T$ returns the same value with grade $1$
    at every honest process.
    This common value must be $v$, since an honest process records
    $v$ in round $T$.
    Thus, every honest process records $v$ and returns it after the loop.

    Therefore, all honest processes return the same value in all cases.
\end{proof}

We are now ready to prove our main theorem:

\begin{theorem}\label{thm:final}
For every collection of prediction vectors,
\cref{alg:ba} solves binary Byzantine Agreement when at most
$t<n/3$ processes are Byzantine.
Let $\predError \leq n^2$ be the total number of incorrect prediction
bits held by honest processes, and let $P=1+\minB$.
Every honest process returns within $\bigO{P}$ communication
rounds, and honest processes send
$\bigO{n^2P}$ messages and $\bigO{n^3+n^2P\log n}$ bits
in total.
\end{theorem}

\begin{proof}
    \emph{Correctness.}
    Since $t<n/3$, all lemmas of this section apply.
    By \cref{lem:termination-simple}, every honest process returns a value.
    By \cref{lem:agreement-simple}, all honest processes return the same value,
    and by \cref{lem:validity-simple}, if all honest processes start with the
    same input $v$, then they all return $v$. Hence \cref{alg:ba} satisfies
    termination, agreement, and validity.

    \emph{Round complexity.}
    \Cref{lem:termination-simple} shows that every honest process
    returns by the end of round
    \[
        R=\min\{t+1,\, f+2,\, k+2\}.
    \]
    Since $f\leq t<n/3$, \cref{lem:k-bound} applies with $\epsilon=1/3$, so
    there is a constant $c\geq 1$ such that $k\leq c\,\predError/n$. Therefore,
    $
        R \leq 2+\min\{f,\, c\,\predError/n\}
          \leq 2+c\min\{f,\, \predError/n\}
          \leq 2+c\,P.
    $
    The classification step takes one communication round, and each round of
    the main loop takes $\bigO{1}$ communication rounds. Hence every honest
    process returns within $\bigO{1}+\bigO{R}=\bigO{P}$ communication rounds.

    \emph{Communication complexity.}
    We count only messages sent by honest processes. 
    With the graded-consensus instantiation specified in the model,
    each invocation takes two communication rounds and each honest
    process sends $\bigO{n}$ messages of $\bigO{1}$ bits each.
    \begin{itemize}
        \item \textbf{Classification.} Every honest process broadcasts its
        $n$-bit prediction vector once. This contributes at most $n^2$
        messages and $n^3$ bits.

        \item \textbf{Main loop.} In each round in which it participates, an
        honest process takes part in three graded-consensus instances
        ($\bigO{n}$ messages of $\bigO{1}$ bits), performs two value
        broadcasts ($2n$ messages of one bit), and broadcasts one id
        in \textsc{Propose} ($n$ messages of $\log n$ bits).
        Thus it sends $\bigO{n}$ messages and $\bigO{n\log n}$ bits per round.
        Since every honest process participates in at most $R$ rounds, the
        main loop contributes $\bigO{n^2R}$ messages and
        $\bigO{n^2R\log n}$ bits.
    \end{itemize}
    Since $R=\bigO{P}$ and $P\geq 1$, honest processes send
    $\bigO{n^2P}$ messages and
    $\bigO{n^3+n^2P\log n}$ bits in total.
\end{proof}

\begin{observation}\label[observation]{obv:bit-complexity}
The initial classification contributes $\bigO{n^3}$ bits,
while the subsequent agreement execution contributes
$\bigO{n^2P\log n}$ bits.
\end{observation}

Exchanging short prefixes of locally predicted honest lists,
with lengths chosen by guess-and-double, removes the initial
$\mathcal{O}(n^3)$ bit cost and reduces total communication to
$\mathcal{O}(n^2P\log n)$ bits. The construction and proof appear
in Appendix~\ref{app:smaller-communication}.

\begin{theorem}
\Cref{alg:ba-with-prefixes} (see Appendix~\ref{app:smaller-communication}) solves binary Byzantine
Agreement when at most $t<n/3$ processes are Byzantine and for any number of 
prediction errors. More specifically,
let $\predError = \bigO{n^2}$ denote the total number of incorrect
prediction bits held by honest processes, and let
$P=1+\minB$.
Every honest process returns within $\bigO{P}$
communication rounds, and honest processes send
$\bigO{n^2P}$ messages and $\bigO{n^2P\log n}$
bits in total.
\end{theorem}

\section{Conclusion}

We presented a deterministic protocol for synchronous binary Byzantine
agreement with classification predictions and optimal resilience
$t<n/3$. Our construction provides a specifically designed prediction-aware Phase King protocol by combining
predetermined and prediction-selected kings. It guarantees correctness
for arbitrarily many prediction errors and terminates in $\bigO{P}$ communication rounds,
where $P=1+\min\{\mathbf{B}/n,f\}$. Thus, accurate predictions accelerate
agreement, while inaccurate predictions preserve the $\bigO{f+1}$
early-stopping guarantee.

In the basic construction complete prediction vectors are exchanged.
We also present a communication-efficient variant, where prefixes of
locally predicted honest lists are exchanged instead, using a guess-and-double strategy
to adapt their lengths as the number of prediction errors is unknown to the processes.
This removes the initial cubic bit cost and achieves
$\bigO{n^2P}$ messages and $\bigO{n^2P\log n}$ bits while preserving
the round bound and resilience.

Our protocol combines state-of-the-art theoretical guarantees (apart from a  $P/\log P$ communication overhead compared to~\cite{dufay2026general}) with practical advantages, due to the avoidance of involved preprocessing (as is the case in~\cite{ben2025byzantine,dufay2026general}). In particular, we believe that our prediction-aware king selection procedure can be of independent interest and can find further applications in learning-augmented consensus. 
In this direction, an interesting open question is whether our techniques can be extended to the case of authenticated protocols.

\newpage
\printbibliography

\newpage
\appendix
\section{Classification with Smaller Communication Cost}
\label{app:smaller-communication}

The classification algorithm (\cref{alg:classify}) requires every process
to broadcast its prediction vector that has size of $\Theta\left(n\right)$,
resulting in $\Theta\left(n^3\right)$ bit complexity. However, the 
king-selection procedure (\cref{alg:propose}) does not require the processes to create
a complete classification vector. Instead, it is sufficient for each honest
process to construct a set $S_i$ such that there exists at least one common
honest process contained in the intersection of those sets: 
$$ \biggl(\bigcap_{i\in\Honest} S_i\biggr) \cap \Honest \neq \emptyset $$

This weaker requirement allows us to avoid sharing the complete prediction vectors.
For each process $p_i$, let
\[
    \predhonest_i
    = \operatorname{sort}\bigl(
        \{j\in\{1,\dots,n\} : a_i[j]=1\}
      \bigr)
\]
be the list of ids that $p_i$ predicts to be honest,
ordered increasingly.
Instead each process can broadcast a shorter prefix of its local list $\predhonest_i$.
An id is then included in $S_i$ only if it appears in sufficiently many
of the received prefixes.

Let $\psize\in\{1,\dots,n\}$ be the common prefix-length.
Each process $p_i$ sends the prefix $\pset_i$ containing the first
$\min\{\psize,|\predhonest_i|\}$ ids  of $\predhonest_i$.
Let $\thres$ be the number of received prefixes in which an id must
appear in order to be included in $S_i$.

For each id $j\in\{1,\dots,n\}$, let $\support_i(j)$
be the number of distinct senders whose proposal list received
by $p_i$ contains $j$, including $p_i$'s own proposal.
Each recipient accepts at most one proposal list per sender
and counts each id at most once in that list.
Define
\[
    S_i=\{j\in\{1,\dots,n\}:\support_i(j)\geq\thres\}.
\]

The intuition is that the total number $\predError$ of prediction errors will
restrict the differences of $\pset_i$'s in a way that there will be a common
honest process $\goodcommon$ that will be ``voted'' by at least $\thres$ honest
processes.  Therefore, all honest processes will include $\goodcommon$ in their set
$S_i$.  Processes do not know $\predError$ but let $\upError \geq
\predError$ be a common upper bound on the total number of incorrect bits held
by honest processes. 

Let $\tol>0$ be the number of prediction errors that must be ``spent'' to
make one honest process unsuitable as the common candidate.  Then for an honest
process $p_j \in \Honest$, define
    $z_j = \left| \left\{i \in \Honest  :  a_i[j] = 0 \right\}\right|$,
i.e. $z_j$ is the number of honest processes that incorrectly predict $p_j$ to
be faulty.

\begin{definition}[\good process]
    An honest process $p_j$ is \good if $z_j < \tol$, where 
    $z_j = \left| \left\{i \in \Honest  :  a_i[j] = 0 \right\}\right|$.
\end{definition}

\begin{lemma}
    \label[lemma]{lem:good}
    Let $\nongoodcount$ denote the number of honest processes that are not
    \good. Then
        $\nongoodcount \leq \left\lfloor\frac{\upError}{\tol}\right\rfloor$.
\end{lemma}

\begin{proof}

    For each honest process $p_j \in \Honest$ that is not \good, we have
    $z_j \geq \tol$. So,
    there are at least $\tol$ prediction errors. The total prediction errors are
    bounded by $\upError$, so $\nongoodcount \cdot \tol \leq \upError \Rightarrow \nongoodcount  \leq
    \lfloor{\frac{\upError}{\tol}}\rfloor$.

\end{proof}

If the number of honest processes that are not \good is $\nongoodcount \geq n-t$ then our
Byzantine Agreement will finish via the predetermined kings. Therefore,
we focus on the ``useful-prediction'' regime, i.e. $\nongoodcount < n -t$.

\begin{observation}
    In the useful-prediction regime $\nongoodcount < n-t$ there exists at least one honest 
    process that is \good. 
\end{observation}

Let $\goodcommon$ denote the honest process with the smallest id that is \good.  We
want to ensure that enough honest (at least $\thres$) processes will broadcast 
$\goodcommon$ so that each honest process $p_i$ will include it in their $S_i$.

\begin{lemma}
    \label[lemma]{lem:common_proposition}
    Suppose $\nongoodcount<n-t$ and $\psize>\nongoodcount$.
    The smallest \good honest id $\goodcommon$ is contained in the
    prefixes of at least
    $
        \left\lceil n-t-\tol-
        \frac{\upError}{\psize-\nongoodcount}\right\rceil
    $
    honest processes.
\end{lemma}

\begin{proof}
    Since $z_{\goodcommon}<\tol$, more than $n-t-\tol$ honest processes
    predict $\goodcommon$ to be honest. Each such process
    $p_h$ sends the first $\psize$ smallest ids that it predicts to be honest.  Suppose
    that $p_h$ does not include $\goodcommon$ in its first $\psize$ locally-honest
    ids.  Since $p_h$ predicts $\goodcommon$ as honest, there must be at least $\psize$
    ids preceding $\goodcommon$ that $p_h$ predicts to be honest.

    By the minimality of $\goodcommon$, every smaller honest id is
    not \good. Thus, by \cref{lem:good}, at most $\nongoodcount$ of these ids can be actually honest. Hence at
    least $\psize - \nongoodcount$ of these ids correspond to Byzantine processes that $p_h$
    incorrectly predicts to be honest. Each such prediction is an incorrect
    prediction bit.  
    
    Consequently, every honest supporter of $\goodcommon$ that fails to broadcast it
    accounts for at least $\psize-\nongoodcount$ prediction errors.    
    The total number of prediction errors is at most $\upError$. Hence, the
    number of such honest processes is at most $\lfloor
    \frac{\upError}{\psize-\nongoodcount} \rfloor.$

    Therefore, the number of honest prefixes containing $\goodcommon$ is
    greater than $n-t-\tol-\upError/(\psize-\nongoodcount)$.
\end{proof}

\begin{lemma}
    \label[lemma]{lem:common_process_sets}
    
    Suppose $\nongoodcount<n-t$, $\psize>\nongoodcount$, and
    $
        0<\thres\leq
        \left\lceil n-t-\tol-
        \frac{\upError}{\psize-\nongoodcount}\right\rceil.
    $
    Then the sets $S_i$ constructed by the honest processes with threshold $\thres$
    satisfy $\left( \bigcap_{i\in \Honest}S_i \right)\cap \Honest \neq\emptyset.$
    In particular, $ \goodcommon\in S_i \ \forall i\in \Honest. $

\end{lemma}

\begin{proof}

    By \cref{lem:common_proposition}, at least $\thres$ honest processes
    broadcast proposal lists containing $\goodcommon$.
    Every honest process receives the proposal of every honest sender.  Therefore, for every honest $p_i$, $
    \support_i(\goodcommon) \ge\thres.  $ By the definition of $S_i$,
    $ \goodcommon\in S_i.$
    Moreover, $\goodcommon$ is honest by construction.  Thus  $\goodcommon \in \left(
    \bigcap_{i\in \Honest}S_i \right)\cap \Honest, $ which proves the claim.    

\end{proof}

For the subsequent king-selection procedure (\cref{alg:propose}), it is also
important to bound the total number of possible candidates appearing in any
honest process's set. Define $$ \candidateunion := \bigcup_{i\in \Honest}S_i. $$ The round
complexity of \cref{alg:propose} is linear in $|\candidateunion|$ as we can deduce from
\cref{lem:progress}.

\begin{lemma}
\label{lem:union-size}

For any threshold $\thres>t$, the union $\candidateunion$ satisfies 
$|\candidateunion|\leq\frac{n\psize}{\thres-t}$.

\end{lemma}

\begin{proof}
    
    Consider any id $j\in \candidateunion$. Then $j\in S_i$ for at least one honest
    process $p_i \in \Honest$.  Therefore, $\support_i(j) \geq \thres$.
    At most $t$ of these proposals can originate from Byzantine processes.  So,
    $j$ must have been proposed by at least $\thres - t$ honest processes.
    Each honest process proposes at most $\psize$ ids. Hence the total 
    number of proposal ids generated by honest processes is at most $n\psize$.
    Every id belonging to $\candidateunion$ accounts for at least $\thres-t$ of these honest
    proposal ids. Thus,
    $
        |\candidateunion|(\thres-t)\leq n\psize.
    $
    Dividing by the positive quantity $\thres-t$ proves the claim.

\end{proof}

\subsection{Choosing $\psize$ and $\tol$}
Let $N = n-2t$ and $q=n-t-\tol-\frac{\upError}{\psize-\upError/\tol}$.
If $\upError/\tol<n-t$, $\psize>\upError/\tol$, and $q>t$, choosing
$\thres=\lceil q\rceil$ satisfies \cref{lem:common_process_sets}.

By \cref{lem:union-size}, the resulting union size $|\candidateunion|$ is bounded by
 $$\sizeobjective(\psize;\tol) = \frac {n\psize} {n-2t-\tol - \frac {\upError} {\psize
- \lfloor\frac{\upError}{\tol}\rfloor}}.$$
For fixed $0<\tol<N$ we want to minimize $\sizeobjective(\psize;\tol)$. 
The $\psize$ that minimizes $\sizeobjective$ is: 
$\psize^*(\tol) = \frac{\upError
\sqrt{N}}{\tol(\sqrt{N} - \sqrt{\tol})}.$
Substituting this into the bound on $\candidateunion$ gives:
$
    |\candidateunion| \leq \frac{n\upError}{\tol(\sqrt{N} - \sqrt{\tol})^2}.
$
Now we optimize for $\tol$. Let $\tolobjective(\tol) = \frac{n\upError}{\tol(\sqrt{N} -
\sqrt{\tol})^2}.$
Then, the $\tol$ that minimizes $\tolobjective$ is
$ \tol^* = \frac{N}{4} $.

Then the prefix length $\psize$ and the threshold $\thres$ become:
\begin{align*}
    \psize^* &= \frac{8\upError}{N} & \thres &= \frac{n}{2}
\end{align*}

\begin{corollary}[Prefix parameters]\label[corollary]{cor:parameters}
Let $N=n-2t$, let $\psize\in\{1,\dots,n\}$, and suppose
$
    \predError\leq\upError\leq\frac{N\psize}{8}.
$
Set $\tol=N/4$ and construct the candidate sets with threshold
$\thres=\lceil n/2\rceil$. Then
$
    \goodcommon\in\bigcap_{i\in\Honest}S_i,
    \
    |\candidateunion|\leq\frac{2n\psize}{N}<6\psize.
$
In particular, the candidate sets share an honest id.
\end{corollary}

\begin{proof}
By \cref{lem:good},
$
    \nongoodcount\leq\frac{\upError}{\tol}
    \leq\frac{\psize}{2}\leq\frac n2<n-t.
$
Thus $\goodcommon$ exists and $\psize>\nongoodcount$. Moreover,
$
    \frac{\upError}{\psize-\nongoodcount}
    \leq\frac{2\upError}{\psize}\leq\frac N4=\tol.
$
The support bound of \cref{lem:common_proposition} is therefore at least
$
    \left\lceil n-t-2\tol\right\rceil
    =\left\lceil\frac n2\right\rceil=\thres.
$
By \cref{lem:common_process_sets}, $\goodcommon$ belongs to every honest
candidate set. Since $\thres-t\geq N/2>0$, \cref{lem:union-size} gives
$|\candidateunion|\leq2n\psize/N<6\psize$, using $t<n/3$.
\end{proof}

Since $N=\Theta(n)$, this gives $\psize=\Theta(1+\upError/n)$ and $|\candidateunion|=\bigO{1+\upError/n}$.
\subsection{Guess-and-double Wrapper}
\label{sec:guess-and-double}

To remove the assumption that processes know an upper bound
$\upError$ on the number of prediction errors, we use a
guess-and-double strategy. Processes execute synchronized phases
with geometrically increasing guesses, which determine both
the prefix lengths used to construct the candidate sets and
the number of iterations executed in each phase.

Phases are indexed by
$\phi=0,\dots,\lceil\log_2(t+1)\rceil$.
Write $N=n-2t$. In phase $\phi$, define
\[
    \psize_\phi=2^\phi,
    \qquad
    \upError_\phi=\frac{N\psize_\phi}{8},
    \qquad
    L_\phi=\left\lceil\frac{2n\psize_\phi}{N}\right\rceil+1.
\]

Once $\upError_\phi\geq\predError$,
the candidate-set bound guarantees a common honest king within
$L_\phi - 1$ iteration.
The additional iteration allows all honest processes to return.
Predetermined kings are used consecutively across phases,
ensuring termination even when all guesses underestimate the
number of prediction errors.

Each phase begins with one communication round in which
processes broadcast prefixes of at most $\psize_\phi$
ids.
In each phase, processes compute support counts using only
the prefixes exchanged in that phase, counting their own
prefix as well.
Processes construct $S_i$ using threshold
$\thres=\lceil n/2\rceil$ and set
$\lead_i=\operatorname{sort}(S_i)$.
The leader list remains fixed within the phase.

Let $r$ be the index of the next predetermined king
at the beginning of the phase.
Each process executes $L_\phi$ iterations in phase $\phi$,
unless it returns earlier.
Each iteration contains the two king steps and final
graded-consensus check of \cref{alg:ba}.
The first king is $p_r$, while the second is selected
by \textsc{Propose} using the current leader list.
Each iteration takes $\bigO{1}$ rounds.

The counter $r$ is initialized once and incremented after
each completed iteration.
Thus, predetermined kings are used consecutively across
phases.
The current value and recorded decision also persist
between phases; only the candidate lists and proposal
positions are reinitialized.
The predetermined kings guarantee progress as a function
of $f$, while the selected kings can accelerate agreement
when the candidate sets contain a common honest id.

A process records a decision when the final graded consensus
returns grade $1$.
It participates in one further iteration.
After iteration $t+1$, every remaining honest process
returns its recorded decision.

\begin{algorithm}[t]
\caption{Byzantine Agreement with Predictions
         (code for process $p_i$)}
\label{alg:ba-with-prefixes}
\begin{algorithmic}[1]
\Procedure{ba-with-prefixes}{$x_i,\preds_i$}
    \State $\predhonest_i\gets
        \operatorname{sort}(\{j:a_i[j]=1\})$
    \State $v_i\gets x_i$, $decided_i\gets\textbf{false}$, $r\gets1$, $N\gets n-2t$

    \For{$\phi=0$ to $\lceil\log_2(t+1)\rceil$}
        \State $\psize_\phi\gets2^\phi$, $\upError_\phi\gets N\psize_\phi/8$, $L_\phi\gets\lceil2n\psize_\phi/N\rceil+1$
        \State $\pset_i\gets\text{first }
            \min\{\psize_\phi,|\predhonest_i|\}
            \text{ ids of }\predhonest_i$
        \State $\Broadcast(\pset_i)$
        \State $S_i\gets
            \{j\in\{1,\dots,n\}:
              \support_i(j)\geq\lceil n/2\rceil\}$

        \State $\lead_i\gets\operatorname{sort}(S_i)$
        \If{$\lead_i=\varnothing$} $\lead_i\gets[i]$
        \EndIf
        \State $\pos_i\gets1$

        \For{$\ell=1$ to $L_\phi$}
            \State $(v_i,g_i)\gets\GradedConsensus(v_i)$
            \State $\Broadcast(v_i)$
            \If{$g_i=0$} $v_i\gets v_r$
                    \Comment{Predetermined king $p_r$}
            \EndIf

            \State $(v_i,g_i)\gets\GradedConsensus(v_i)$
            \State $(\king_i,\pos_i)\gets
                \textsc{Propose}(\lead_i,\pos_i)$
            \State $\Broadcast(v_i)$
            \If{$g_i=0$} $v_i\gets v_{\king_i}$
                    \Comment{Selected king}
            \EndIf

            \State $(v_i,g_i)\gets\GradedConsensus(v_i)$

            \If{$decided_i$}
                \State \Return $decision_i$
            \EndIf

            \If{$g_i=1$}
                \State $decision_i\gets v_i$
                \State $decided_i\gets\textbf{true}$
            \EndIf
            \If{$r=t+1$} \Return $decision_i$
            \EndIf
            \State $r\gets r+1$
        \EndFor
    \EndFor

    \State \Return $decision_i$
\EndProcedure
\end{algorithmic}
\end{algorithm}

\begin{theorem}\label{thm:prefix-ba}
\Cref{alg:ba-with-prefixes} solves binary Byzantine
Agreement when at most $t<n/3$ processes are Byzantine and for any number of 
prediction errors. More specifically,
let $\predError = \bigO{n^2}$ denote the total number of incorrect
prediction bits held by honest processes, and let
$P=1+\minB$.
Every honest process returns within $\bigO{P}$
communication rounds, and honest processes send
$\bigO{n^2P}$ messages and $\bigO{n^2P\log n}$
bits in total.
\end{theorem}

\begin{proof}
Processes retain their current values
and recorded decisions across phases. Therefore, the preservation,
decision-propagation, and agreement arguments (\cref{lem:preservation-simple,lem:decision,lem:agreement-simple})
for \cref{alg:ba} remain valid. The phase range $\psize_\phi \leq n$
provides at least $t+1$ iterations, so one of the first $f+1$ predetermined king
will be honest. Thus every honest process records a decision and
returns by iteration
$
    R\leq\min\{t+1,f+2\}.
$
Agreement and validity follow as for \cref{alg:ba}.

To prove the prediction-dependent bound, consider a phase with $\upError_\phi\geq\predError$.
If an honest process has already recorded a decision,
propagation ensures termination within two further iterations.
Otherwise, all honest processes participate in the prefix
exchange. Let $K$ be the size of the union of the honest candidate sets. 
By \cref{cor:parameters}, their candidate sets
share a common honest id $\goodcommon$:
\[
    \goodcommon\in\bigcap_{i\in\Honest}S_i \quad \text{ and } \quad
    K:=|\candidateunion|\leq\frac{2n\psize_\phi}{N}.
\]
There are at most $K-1$ proposed ids below the
shared honest id $\goodcommon$. The progress and convergence arguments
of \cref{lem:progress,lem:convergence} therefore give a common
honest king within $K$ iterations. If a first decision occurs in iteration
$a<K$, all honest processes return by iteration $a+2\leq K+1$.
Otherwise, the common honest king makes everyone record a
decision by iteration $K$ and return in the next iteration.
Since $K+1\leq L_\phi$, all honest processes return by the
end of this phase. If the global iteration counter reaches
$t+1$ earlier, every remaining honest process returns then.

In each phase the number of iterations are $\L_\phi = 2 \frac{n}{N} 2^\phi + 1$.
Since $n > 3t$, then $1 \leq \frac{n}{N} < 3$. Therefore, 
$
    2\cdot2^\phi\leq L_\phi\leq 6\cdot2^\phi + 1 \leq 7 \cdot 2^\phi.
$

So the number of iterations until a phase $\phi$ are:
\[
    \sum_{\psi=0}^{\phi}L_\psi
    \leq7(2^{\phi+1}-1)=\bigO{2^\phi}.
\]

Let $\phi^\ast$ be the first phase where $\upError_{\phi^\ast} \geq \predError$, which means
$N2^{\phi^\ast}/8\geq\predError$.
If $\phi^\ast=0$, then $2^{\phi^\ast}=1$.
Otherwise, the upper error bound for the previous phase would be less than $\upError$, i.e. 
$\upError_{\phi^\ast - 1} < \predError$. Therefore,
$
    \frac{N2^{\phi^*-1}}8<\predError,
    \text{ hence }
    2^{\phi^*}<\frac{16\predError}{N}.
$
Consequently,
$$
    2^{\phi^*}=\bigO{1+\predError/n}.
$$

Let $R$ denote the number of iterations until every honest process has returned, and $\Phi = \lceil\log(t+1)\rceil$ be the last scheduled 
phase.

If $\phi^\ast\leq\Phi$, all honest processes return by the end
of phase $\phi^\ast$, or earlier. The geometric sum above gives
$
    R=\bigO{2^{\phi^*}}=\bigO{1+\predError/n}.
$
If $\phi^\ast>\Phi$, in each phase the upper error bound is not greater than the real number of errors. In
particular,
$
    \predError>\frac{N2^\Phi}{8}
    \geq\frac{N(t+1)}8.
$
In this case, the honest nodes terminates through the predetermined kings in rounds
$
    R\leq t+1<\frac{8\predError}{N}
    =\bigO{\predError/n}.
$

Therefore $R=\bigO{1+\predError/n}$ in both cases.
Independently, predetermined kings give $R\leq f+2$.
Combining the two bounds,
\[
    R=\bigO{\min\{f+1,1+\predError/n\}}
     =\bigO{1+\min\{f,\predError/n\}}
     =\bigO{P}.
\]

\emph{Communication complexity.}
In each phase $\phi$ the processes exchange a prefix of their prediction vectors, that each one has size $m_\phi = 2^\phi$.
This gives $\bigO{n^2}$ messages for each round and $\bigO{n^2 \cdot 2^\phi \log n}$ bits.

Let $q$ be the last phase reached. Then for the prefixes sizes holds 
$
    \sum_{\phi=0}^{q}\psize_\phi
    =2^{q+1}-1\leq R.
$

Therefore, for $q$ phases for the prefixes exchanges the total number of bits are:
\[
    \quad \bigO{n^2\log n\sum_{\phi=0}^{q}\psize_\phi}
    =\bigO{n^2R\log n}
\]

Each iteration also uses $\bigO{n^2}$ messages and $\bigO{n^2 \log{n}}$ bits from graded-consensus and propose.
The total number of iterations are $R = \bigO{P}$. So the total nubmer of messages and bits respctively are:
\[
    \bigO{n^2 P} \text{ messages }, \qquad \bigO{n^2 P \log{n}} \text{ bits.}
\]
\end{proof}
\section{Notation Table} \label{app:notation}

The table~\ref{tab:notation} below summarizes the notation used throughout the paper.

\begin{table}[H]
\centering
\small
\renewcommand{\arraystretch}{1.15}
\begin{tabular}{@{}p{0.27\linewidth}p{0.67\linewidth}@{}}
\hline
\textbf{Symbol} & \textbf{Meaning} \\
\hline
$n$ & Number of processes. \\
$t$ & Upper bound on the number of Byzantine processes. \\
$f$ & Actual number of Byzantine processes. \\
$\Honest,\Faulty$
    & Sets of honest and Byzantine processes, respectively. \\
$\preds_i$
    & Raw prediction vector held by process $p_i$. \\
$\clas_i$
    & Classification vector computed by process $p_i$. \\
$\Honest_i$
    & Sets of ids classified as honest according to $\clas_i$. \\
$\lead_i$
    & the ids of $\Honest_i$ (or of $S_i$ in
      \cref{app:smaller-communication}) in increasing order. \\
$\predhonest_i$
    & Increasingly ordered list of ids predicted to be honest
      according to $\preds_i$. \\
$\predError$
    & Total number of incorrect prediction bits held by honest processes. \\
$\upError$
    & Common upper bound on $\predError$. \\
$P$
    & Round parameter $P=1+\min\{\predError/n,f\}$. \\
$k$
    & Number of ids misclassified by at least one honest process. \\
$\pos_i(r)$
    & Position in $\lead_i$ of the id proposed by $p_i$ in round $r$. \\
$q_i(r)$
    & id proposed by process $p_i$ in round $r$. \\
$s_i(r)$
    & King selected by $p_i$ in round $r$. \\
$\tol$
    & Error threshold defining an $\tol$-good honest process. \\
$V_i^r(j)$
    & Number of votes for id $j$ received by $p_i$ in round $r$. \\
$U_i^r(s),\widehat U_i^r(s)$
    & Number of votes received by $p_i$ in round $r$ for ids
      at least $s$, counting all ids or only those in
      $\lead_i$, respectively. \\
$\candidateunion$
    & Union of the candidate sets of honest processes:
      $\candidateunion=\bigcup_{i\in\Honest}S_i$. \\
$\psize$
    & Maximum number of ids included in a proposal prefix. \\
$\thres$
    & Support threshold for including an id in $S_i$. \\
\hline
\end{tabular}
\caption{Notation used throughout the paper.}
\label{tab:notation}
\end{table}

\end{document}